\documentclass[aps,prb,twocolumn,superscriptaddress,10pt]{revtex4-1}

\usepackage{amssymb,amsmath}
\usepackage{graphicx,color}

\begin{document}

\title{The self-actuating InAs nanowire-based nanoelectromechanical Josephson junction}

\author{Andrey V. Kretinin}
\email{andrey.kretinin@manchester.ac.uk}
\affiliation{Weizmann Institute of Science, Condensed Matter Physics Department, Rehovot, Israel}
\author{Anindya Das}
\affiliation{Weizmann Institute of Science, Condensed Matter Physics Department, Rehovot, Israel}
\author{Hadas Shtrikman}
\affiliation{Weizmann Institute of Science, Condensed Matter Physics Department, Rehovot, Israel}

\maketitle

\textbf{Half a century ago Brian Josephson made a series of striking predictions related to a tunnelling barrier sandwiched between two superconductors.\cite{Josephson1962} One particular prediction, later became known as the a.c.-Josephson effect, said that under a finite d.c. bias the tunnelling current contains an a.c. supercurrent component, oscillating at microwave frequency. This prediction was experimentally verified through observation of the junction current-voltage characteristics modified by the interaction with the electromagnetic radiation,\cite{Shapiro1963,Coon1965} and direct detection of the microwave radiation emitted from the junction itself.\cite{Yanson1965} It had also been established that the behaviour of the d.c. current-voltage characteristic is determined by the high-frequency dynamics of the Josephson junction,\cite{McCumber1968} and can be used to study various systems such as atoms in the electron-spin resonance,\cite{Baberschke1984} optical phonons in high-$T_{\rm c}$ superconductors,\cite{Helm1997} and vibrating Nb molecules.\cite{Marchenkov2007} Here we present an InAs nanowire Josephson junction device,\cite{Doh2005,Xiang2006} where a vibrating nanowire weak link plays the role of a nanoelectromechanical resonator. The flow of the a.c.-Josephson current through the junction enabled actuation and detection of vibrational modes of the resonator by means of simple d.c. transport measurements.}

There are two regimes a Josephson junction can be operated in.\cite{Josephson1962} The first, the so-called d.c.-Josephson effect, is characterised by a d.c.~superconducting current $I_{\rm s} = I_{\rm c}\sin\left(\Delta\phi\right)$, where $I_{\rm c}$ is the critical current, and $\Delta\phi$ is the superconducting wave functions phase difference. In the second regime corresponding to the a.c.-Josephson effect the tunnelling current exceeds the value of $I_{\rm c}$, and consists of two components. Apart from a normal dissipative current manifested by the voltage drop $V$, the total current contains the oscillating superconductive part $I_{\rm s} = I_{\rm c}\sin\left(\omega_{\rm j}t + \phi_{0}\right)$, where $\phi_{0}$ is the initial phase difference. In this case the phase difference increases linearly in time $t$ with the rate set by the d.c. voltage as $\omega_{\rm j} = (2e/\hbar)V$.

Because of the high frequency of the a.c.-Josephson current ($2e/h\approx$~483.6~MHz/$\mu$V), the shape of the Josephson junction current-voltage ($I-V$) characteristics is determined by its electrical impedance.\cite{Coon1965,McCumber1968} One of the most vivid demonstrations of this fact was the experiment of Coon and Fiske,\cite{Coon1965} where the $I-V$ characteristics of the Josephson junction coupled to the electromagnetic resonator, exhibited constant-voltage step features (self-induced Shapiro steps). In this experiment the steps occurred when the voltage $V$ was such that the a.c.-Josephson oscillations were in resonance, namely, at voltage $V_{n} = \hbar\,\omega_{n}/2e$, where $\omega_{n}$ is the frequency of the $n$-th resonator mode. This work illustrated how a Josephson junction coupled to an external resonant system can be simultaneously used as a source and detector of the microwave radiation. This remarkable property allowed the use of a Josephson junction as an ultra-small power spectrometer to study atomic and vibrational spectra.\cite{Baberschke1984,Helm1997,Marchenkov2007}

\begin{figure*}[!ht]
  \includegraphics{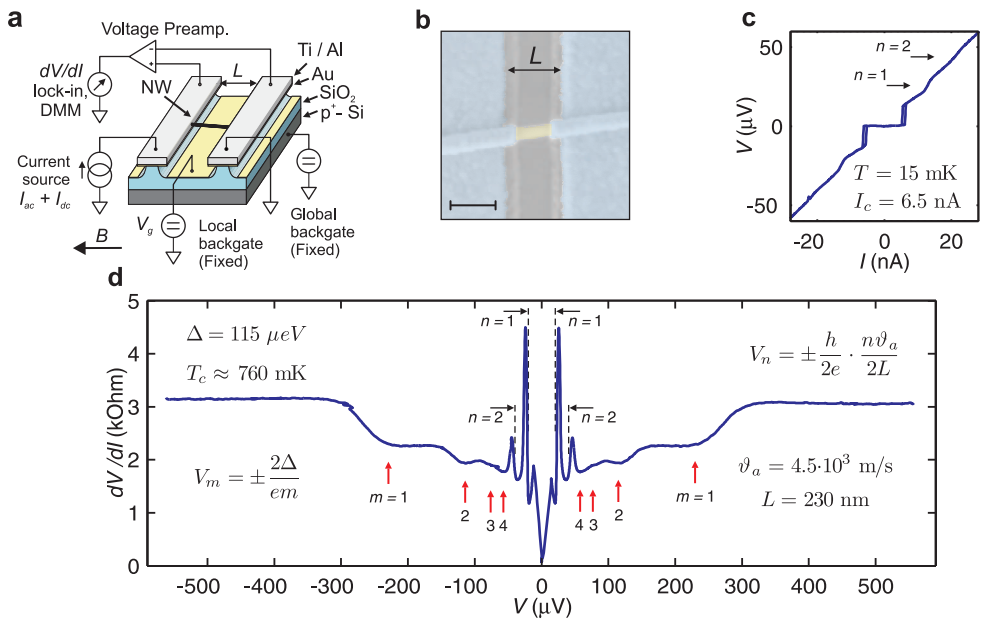}\\
  \caption{\textbf{The suspended InAs nanowire-based Josephson junction and its $I-V$ characteristics.} \textbf{(a)} A schematic representation of the suspended nanowire device and the experimental set-up. The InAs nanowire (NW) is suspended over a trench predefined in SiO$_{2}$, and clamped by two Ti/Al contacts. An in-plane magnetic field $B$ was applied during the experiment, and it was only few degrees off the nanowire axis. \textbf{(b)} The top view of a typical device obtained by a scanning electron microscope. The scale bar is 200~nm. The fake colours indicate different parts of the device. Blue is the Ti/Al contacts, yellow is the nanowire, and gray is the local backgate below the nanowire. \textbf{(c)} The $I-V$ characteristics of the nanowire Josephson junction taken at $V_{\rm g}=$~12.3~V. \textbf{(d)} The differential resistance $dV/dI$ as a function of the voltage $V$ taken at the same $V_{\rm g}$ as in \textbf{(c)}. The red arrows show first four orders of MAR at voltage $V_{\rm m}$. The position of the first MAR signature ($m=$~1) gives the superconducting gap $2\Delta\approx$~230~$\mu$eV and the estimate for $T_{\rm c}=2\Delta/(3.52\,k_{\rm B})\approx$~760~mK. The black arrows mark the resistance resonances caused by the acoustic wave actuated by Josephson oscillations at voltage $V_{\rm n}$.}
  \label{Fig1}
\end{figure*}

Recently, the combination of the Josephson junction and nanoelectromechanical system\cite{Ekinci2005} (NEMS) has attracted increasing attention. It was proposed to use the nanomechanical resonator to store the Josephson junction phase qubit states,\cite{Cleland2004,O'Connell2010} and to build a coherent source of microwave radiation based on an array of Josephson junctions.\cite{Trees2005} In addition, a vibration-modulated Josephson junction has been predicted to exhibit self-induced Shapiro steps in the $I-V$ characteristics at a frequency of mechanical resonance,\cite{Zhu2006} and the weak-link made of a doubly-clamped nanowire can be used to pump the supercurrent into nanowire vibrations.\cite{Sonne2008} In all these aspects NEMS act similarly to an external microwave resonator coupled to a Josephson junction.

\begin{figure*}[!ht]
  \includegraphics[width=0.8\textwidth]{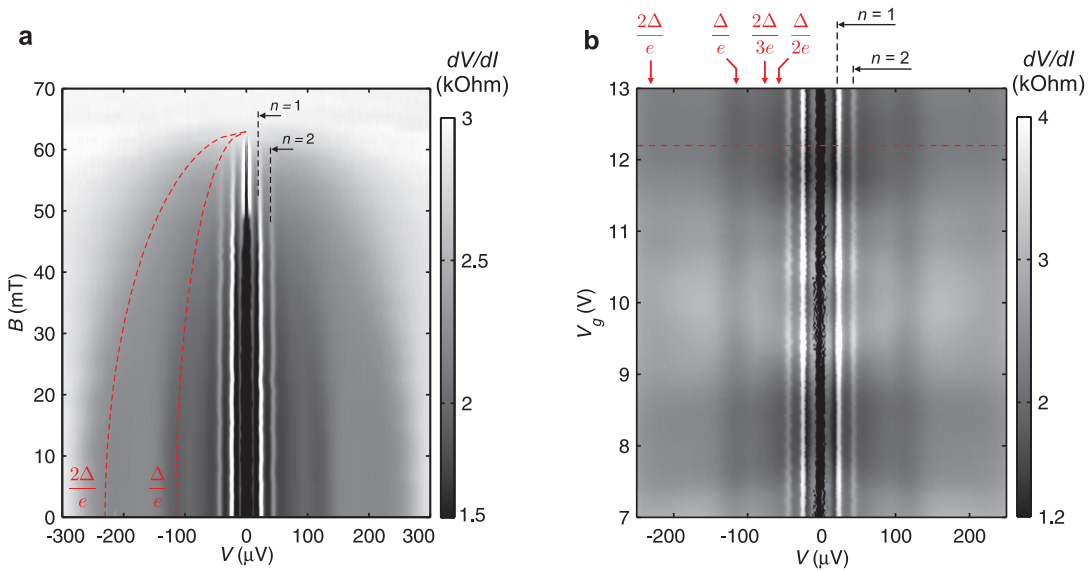}\\
  \caption{\textbf{The magnetic field- and gate voltage-independent position of the resistance resonances.} \textbf{(a)} The gray-scale plot of $dV/dI$ made in the $V-B$ plane at $T = T_{\rm base}$. The red dashed curves show the position of the first two orders of MAR expected from the Landau-Ginzburg theory,\cite{Douglass1961} $\Delta(B)/\Delta(0) = \sqrt{1-(B/B_{\rm c})^2}$, where $B_{\rm c}=$~63~mT is the critical field. The black arrows mark the magnetic field-independent position of the resistance resonances. The position of the resistance resonances stays unchanged while the peak amplitude decreases while $B$ approaches its critical values $B_{\rm c}$. \textbf{(b)} The gray-scale plot of $dV/dI$ made in the $V-V_{\rm g}$ plane at the same temperature as in \textbf{(a)}. The red arrows point at the first four MAR signatures seen in Fig.~\ref{Fig1}d. The black arrows show the gate voltage-independent resistance resonances. The horizontal red dashed line marks the $dV/dI$ trace shown in Fig.~\ref{Fig1}d taken at $V_{\rm g}=$~12.3~V. The changing shades of gray quasi-periodic in $V_{\rm g}$ are due to the Fabry-P\'{e}rot quasiparticle conductance oscillations.\cite{Kretinin2010}}
  \label{Fig2}
\end{figure*}

Here we report observation of the a.c.-Josephson oscillations being in resonance with the longitudinal acoustic wave of an InAs nanowire. Using a previously developed technique\cite{Kretinin2010} we fabricated a new type of nanowire Josephson junction device\cite{Doh2005,Xiang2006} (Figs.~\ref{Fig1}a and ~\ref{Fig1}b). In our field-effect type device the 240~nm-long nanowire weak-link is suspended in vacuum and clamped by two superconducting Ti/Al contacts, making the nanowire-based NEMS\cite{Solanki2010} an integral part of the Josephson junction. The transport experiments with the current-driven Josephson junction described here were carried out inside the dilution refrigerator with $T_{\rm base}\approx$~15~mK. All measurements were taken at a fixed voltage on the local and global backgates. Figure~\ref{Fig1}c shows the experimental $I-V$ curve typical of an underdumped Josephson junction; a non-dissipative branch limited by the critical current $I_{\rm c}\approx$~6.5~nA, some hysteresis at around $I_{\rm c}$, and the resistive part, with the well-pronounced step-like features. To look into the details of the $I-V$ curve we measured the differential resistance $dV/dI$ as a function of applied voltage $V$ (Fig.~\ref{Fig1}d). At the subgap voltages ($e|V|<2\Delta$, here $2\Delta$ is the superconducting gap in the Al contacts) the differential resistance reveals an overall decrease with some dips originating from the excess current of the so-called Multiple Andreev Reflections\cite{Octavio1983} (MAR). The dips in the resistance occur when the superconducting gap $2\Delta$ is an integer multiple of the applied bias, which defines the position of the MAR signature as $V_{m} = \pm2\Delta/em$, where $m$ is integer. In our experiment we clearly observed MAR signatures with $m$ up to four (red arrows in Fig.~\ref{Fig1}d). However, at lower biases, the subgap features become sharp well-recognised asymmetric resistance resonances (black arrows in Fig.~\ref{Fig1}d), corresponding to the steps on the $I-V$ curve (Fig.~\ref{Fig1}c). The distinct nature of these resonances becomes apparent from the gray-scale plot of $dV/dI$ measured at different magnetic fields shown in Fig.~\ref{Fig2}a. The position of the MAR signatures is tightly bound to the value of the magnetic field- and temperature-dependent gap $2\Delta$, and shifts toward $V=$~0 as the magnetic field $B$, or temperature $T$, approach their critical values. As seen from Fig.~\ref{Fig2}a, the position of the MAR signatures for $m=$~1,~2 follow the prediction of the Landau-Ginzburg theory\cite{Douglass1961} (red dashed curves). Whereas, the position of the resistance resonances (black arrows) remains unchanged, and the resonances simply vanish for $B\geqslant B_{\rm c}$. Qualitatively the same behaviour is observed for $dV/dI$ measured at different temperatures. Moreover, the position of the resonances is independent of the backgate voltage $V_{\rm g}$ (Fig.~\ref{Fig2}b), which rules out the superconducting bound states\cite{Gennes1963,Ingerman2001} as possible explanation for there origin.

Such a peculiar behaviour indicates that these features are due to the self-induced Shapiro steps of the junction coupled to an external resonator.\cite{Dayem1966} The first resonance appears at around $V\approx$~20~$\mu$V, which corresponds to an a.c.-Josephson frequency of about 10~GHz. In our experiment there was no deliberately designed microwave resonator coupled to the nanowire junction. The possibility of a microwave cavity unintentionally formed inside our cryogenic setup was excluded by reproducing the same result in a different cryogenic system. The distributed parameter resonators, such as coplanar striplines, are ruled out by the fact that the dimensions of the device including the leads do not exceed 1~mm, and it is much smaller than the radiation wavelength ($\sim$~3~cm). We also considered a lumped $LC$ circuit, where the capacitance is defined by the area of the device bonding pads ($\sim$~3~pF), and the inductance is defined by the device leads ($\sim$~1~nH). The estimated resonant frequency ($\sim$~3~GHz) appears to be well below the measured one.

The important clue to the understanding of the observed resistance resonances lies in the fact that the nanowire junction is suspended and its possible mechanical oscillations are not dumped by the substrate. Indeed, a quick estimation shows that the a.c.-Josephson frequency of the resonances turn out to be close to the frequency of the first two longitudinal acoustic modes of the doubly-clamped InAs beam\cite{LandauLifshitzBook} $f_{\rm n} = (\vartheta_{\rm a}/2L)\times n \approx 9.4$ GHz$\times n$, where $n$ - is integer, $\vartheta_{\rm a}=4.5\times10^{3}$~m/s is the speed of sound in wurtzite InAs nanowire along the $\langle111\rangle$ direction,\cite{Mariager2010} and $L=$~240~nm is the length of the beam. This suggests that the resistance resonances are caused by interaction of the a.c.-Josephson current with the acoustic modes of the nanowire. Since the mode frequency is defined only by the nanowire length the position of the resonances caused by them should be independent of the values of $\Delta$ and $V_{\rm g}$, which is confirmed by our observations.

To further understand this phenomenon we used a simple model of the Josephson junction coupled to an InAs nanowire-based Bulk Acoustic Resonator\cite{RosenbaumBook} (BAR).

\begin{figure*}[!ht]
  \includegraphics[width=0.8\textwidth]{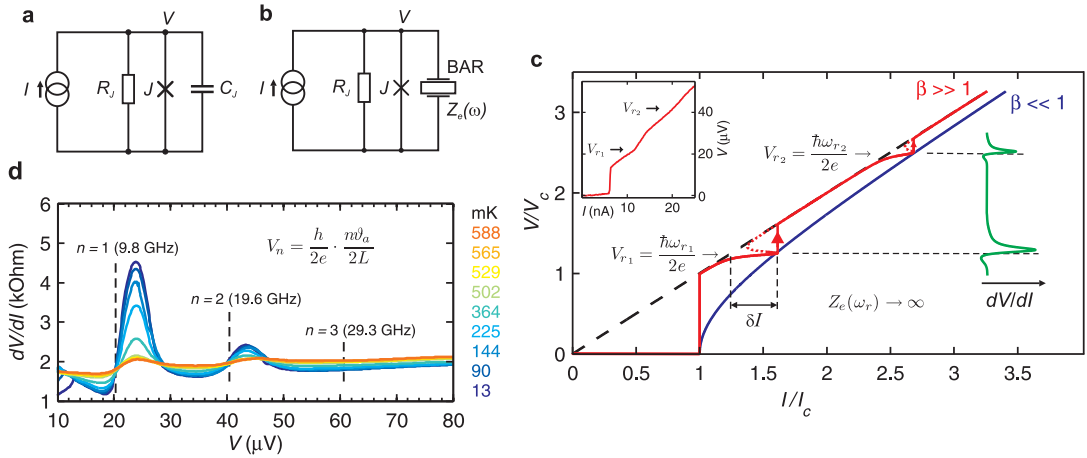}\\
  \caption{\textbf{The RSJ-model and $I-V$ characteristics of the Josephson junction coupled to the Bulk Acoustic Resonator.} \textbf{(a)} The RSJ-model circuit diagram of a typical Josephson junction. \textbf{(b)} The RSJ-model of the Josephson junction coupled to the BAR. \textbf{(c)} The schematic plot of the Josephson junction $I-V$ characteristics. If the capacitance $C_{\rm j}$ is large ($\beta\gg$~1) the a.c.-Josephson current is shunted by $C_{\rm j}$, and its time-averaged contribution to the $I-V$ characteristics is zero. Thus for $I>I_{\rm c}$ the $I-V$ characteristics is of a normal conductor $V = IR_{\rm n}$  (black dashed line). In the opposite case of small $C_{\rm j}$ ($\beta\ll1$) the alternating current is forced to flow through $R_{\rm n}$ causing the a.c.-Josephson current to be frequency-modulated. The time-average of the frequency-modulated a.c. current is non-zero, which makes the total $I-V$ characteristics parabolic\cite{McCumber1968} $V/V_{\rm c}=\sqrt{(I/I_{\rm c})^2-1}$, where $V_{\rm c}=R_{\rm n}I_{\rm c}$ (blue curve). The $I-V$ characteristics of the underdumped Josephson junction coupled to the BAR is shown by the red curve. When the voltage $V=V_{\rm r}$ the Josephson oscillations actuate the acoustic resonance in BAR ($Z_{\rm{e}}(\omega)\rightarrow\infty$), which forces the junction to cross from the underdumped to the overdumped regime, and results in the current peak $\delta I$. For simplicity we assumed that the resonance in $Z_{\rm e}$ has a Lorentzian lineshape. In the case of the current-driven junction, the current peak takes a step-like shape with a multi-valued region showed by the red dotted curve. The green curve on the right-hand side is the corresponding $dV/dI$ as a function of $V$. \textbf{(Inset)} The actual $I-V$ curve measured at $T=T_{\rm {base}}$ and $B=$~0. \textbf{(d)} The zoomed-in plot of the $dV/dI$ resonance taken at different temperatures and zero magnetic field. Identically to the magnetic field dependence (Fig.~\ref{Fig2}a), the position of the resonances is independent of the temperature, which rules out their MAR-related origin. We used the interception point of curves taken at different $T$ to identify the resonant voltage $V_{\rm n}$. The vertical dashed lines designate the voltage $V_{\rm n}$ which corresponds to the first three longitudinal harmonics expected for the doubly-clamped InAs beam with length $L=$~230~nm and speed of sound $\vartheta_{\rm a}=~4.5\cdot10^3$~m/s. The respective frequencies are given in brackets.}
  \label{Fig3}
\end{figure*}

The junction is described by the RSJ-model\cite{McCumber1968} (Fig.~\ref{Fig3}a), where an ideal junction $J$ is connected in parallel to the normal state resistance $R_{\rm n}$, and the total junction capacitance $C_{\rm j}$. As mentioned before, the shape of the $I-V$ characteristics is governed by the high-frequency properties of the junction, namely by the relative impedance of $R_{\rm n}$ and $C_{\rm j}$, and characterised by the parameter $\beta=\omega_{\rm c}R_{\rm n}C_{\rm j}$,\cite{McCumber1968} where $\omega_{\rm c} = (2e/\hbar)I_{\rm c}R_{\rm n}$ is the lowest frequency of the Josephson oscillations. If $C_{\rm j}$ is large ($\beta>>$~1) the junction is `underdumped' and the $I-V$ curve is linear. In the opposite limit of small $C_{\rm j}$ ($\beta<<$~1), the junction is `overdumped' and the $I-V$ curve is parabolic. The experimental values of the critical current ($I_{\rm c}=$~6.5~nA) and the normal-state resistance ($R_{\rm n}\sim$~2~kOhm) gave us the lowest value of the Josephson oscillations frequency $\omega_{\rm c}/2\pi\sim$~6.3 GHz, and $\beta\approx$~30 (with $C_{\rm j}\sim3$~pF), indicating that the junction is `underdumped'.

In our device capacitance $C_{\rm j}$ is a result of the parallel connection of the leads capacitance and the piezoelectric capacitor formed by the wurtzite InAs nanowire.\cite{Xin2007} In the case of a piezoelectric capacitor the external a.c. electric field excites acoustic waves inside the dielectric media and vice versa. The process of energy conversion from the electrical to the mechanical domain profoundly affects the capacitor impedance, and is the basis of the BAR operation. According to the one-dimensional Mason model\cite{RosenbaumBook} the total effective impedance $Z_{\rm e}(\omega)$ of a piezoelectric capacitor is
\begin{equation}
Z_{\rm e}(\omega) = \frac{1}{j\omega C_{\rm j}}\left[1-k_{\rm t}^2\cdot\frac{\tan\left(\frac{\omega L}{2\vartheta_{\rm a}}\right)}{\frac{\omega L}{2\vartheta_{\rm a}}}\right],
\label{Z_eff}
\end{equation}
where $\omega$ is the angular frequency, $k_{\rm t}^2$ is the piezoelectric coupling constant, $\vartheta_{\rm a}$ is the speed of sound in the piezoelectric media, and $L$ is the thickness of the piezoelectric material. It is clear from Eq.~(\ref{Z_eff}) that the longitudinal acoustic standing wave with $\omega_{\rm r} = (\pi/L)\vartheta_{\rm a}$ causes the total impedance to diverge ($Z_{\rm e}(\omega_{\rm r})\rightarrow\infty$).

Now we assume that instead of $C_{\rm j}$ the Josephson junction dynamics is defined by the impedance $Z_{\rm e}$ of the InAs nanowire BAR (Fig.~\ref{Fig3}b). While the BAR is off resonance, the $I-V$ curve follows the one expected for the underdumped regime (blue curve, Fig.~\ref{Fig3}c). When the voltage $V$ is such that the Josephson oscillations are in resonance with the BAR its impedance increases, forcing the Josephson junction to cross from the underdumped to the overdumped regime (red curve, Fig.~\ref{Fig3}c). This crossing results in a cusp-like increase of the d.c.-current through the junction $\delta I$ with the maximum at $V_{\rm r} = \hbar\,\omega_{\rm r}/2e$, where $\omega_{\rm r}$ is the mechanical resonance frequency. In the case of the current-driven Josephson junction the current peak on the $I-V$ curve becomes a step-like feature and corresponds to the asymmetric resonance in $dV/dI$ (green curve, Fig.~\ref{Fig3}c).

Using this model, we determined the resonant frequency of the nanowire BAR actuated by the Josephson oscillations. Two resistance peaks shown in Fig.~\ref{Fig3}d correspond to the resonant coupling of the first two longitudinal acoustic modes of the doubly-clamped InAs nanowire,\cite{LandauLifshitzBook} with onset at the voltage determined as
\begin{equation*}
V_{\rm n} = \pm\frac{\hbar\,\omega_{\rm r}}{2e} = \pm\frac{h}{2e}\cdot\frac{n\,\vartheta_{\rm a}}{2L^{*}},
\end{equation*}
where $n =$~1,~2 and $L^{*}=$~230~nm is the length of the acoustic resonator ($\vartheta_{\rm a}=4.5\times10^{3}$~m/s). Note, the length of the resonator $L^{*}$ found here is close to the physical length of the nanowire Josephson junction (240~nm). Despite the fact that the resonance in $Z_{\rm e}$ requires a non-zero spatially averaged strain,\cite{RosenbaumBook} which forbids the appearance of even harmonics, we have observed the second harmonic as well. The reason for that is the slightly asymmetric geometry of our device.

The piezoelectric coupling of the Josephson oscillations to the acoustic mode of a crystal resonator has been theoretically studied for the case of an array of Josephson junctions.\cite{Trees2005} From this point of view our experimental realisation can be seen as a special case of a single weakly coupled junction. Other possible coupling mechanisms such as the magnetomotive force,\cite{Sonne2008} and modulation of the junction dimensions\cite{Zhu2006} are ruled out since the resonance is observed at $B=$~0, and the junction length $L$ is fixed. It is worth mentioning that the strong coupling between the longitudinal vibrations and single charge tunneling by means of electron-vibron coupling\cite{Mariani2009} has been observed in suspended carbon nanotubes.\cite{Sapmaz2006,Leturcq2009} However, the underlying physical principals responsible for the electron-vibron coupling is entirely different from the one studied in this work.

In conclusion, we fabricated an InAs nanowire nanoelectromechanical Josephson junction, and observed the resonant coupling of Josephson oscillations to the longitudinal acoustic modes of the nanowire. The resonant coupling reveals itself as voltage steps in the junction $I-V$ characteristics occurring when the Josephson oscillations are in resonance with the acoustic modes. We showed that the Josephson junction coupled to the NEMS can be used as a self-actuating probe to study vibrational states. The advantage of this tool is in its simplicity, the NEMS is probed by routine d.c.-measurements, with no need for external high-frequency equipment.\cite{Ekinci2005,Solanki2010,O'Connell2010} Also, our experiment demonstrates the possibility of building a nano-scale source of coherent microwave radiation based on an array of Josephson junctions coupled to a high-$Q$ piezoelectric resonator.\cite{Trees2005}

\textbf{METHODS} -- The high-quality InAs nanowires used in this work have been synthesised by the Molecular-Beam Epitaxy in Riber 32 solid source system by the Au-assisted Vapour-Liquid-Solid (VLS) method. The nanowires were grown along $\langle111\rangle$ direction on a (011)-oriented InAs substrate at 400 $^{\rm{o}}$C and the group V/III ratio of 100. All nanowires have a pure wurtzite crystal structure with only one stacking fault per several nanowires.\cite{Kretinin2010}

The basis of the suspended nanowire Josephson junction was a p$^{+}-$Si wafer with 300~nm thick SiO$_2$ thermally grown on top. First, an array of 250~nm~$\times$~20~$\mu$m trenches was chemically etched in the SiO$_2$ to depth of about 150~nm. This was followed by evaporation of the local metal backgate on the bottom of the trenches. Then nanowires were randomly distributed across the predefined trenches so that a small segment of a nanowire is suspended over the trench. E-beam lithography was employed to pattern Ti/Al contacts (5~nm/100~nm) on top of the parts of the nanowire supported by the banks of the trench, leaving the suspended segment untouched (Fig~\ref{Fig1}a). To ensure high contact transparency the nanowire native oxide was removed with the 0.2$\%$ solution of (NH$_4$)$_2$S$_x$ ($x=$~1.5~M) prior to evaporation of the contacts. The nanowire junction was about 240~nm long with a diameter of about 50~nm (Fig~\ref{Fig1}b). The experimental data obtained from another device is given in the Supplementary Information.

All transport measurements were carried out inside a $^3$He~-$^4$He dilution refrigerator with heavily filtered signal lines and $T_{\rm{base}}\approx$~15~mK. The $I-V$ characteristics and differential resistance as a function of d.c. voltage were measured in a quasi four-terminal constant current configuration. The voltage drop across the device was amplified with a low-noise voltage preamplifier and registered using a high-speed digital voltmeter or lock-in amplifier (Fig.~\ref{Fig1}a). The a.c. excitation current $I_{\rm{ac}}$ was set to be below 400~pArms, so that the excitation voltage drop across the sample was not greater than $k_{\rm B}T$.

\textbf{ACKNOWLEDGMENTS} -- Authors would like to acknowledge Moty Heiblum for making this project possible, and for critical remarks and comments made during the work. The authors are grateful to Eli Zeldov and David Stroud for the enlightening discussions. We also would like to thank Denis Vasyukov, Yuval Oreg, Alex Palevskii for sharing their knowledge of superconductivity, Yunchul Chung for technical ideas, Diana Mahalu for the expertise in e-beam lithography and Ronit Popovitz-Biro for professional TEM study of the InAs nanowires. This work was partially supported by the EU FP6 Program Grant No.~506095, by the Israeli Science Foundation Grant No.~530-08, and Israeli Ministry of Science Grant No. 3-66799.

\end{document}


\title{Supplementary Information for\\The self-actuating InAs nanowire-based nanoelectromechanical Josephson junction}

\author{Andrey V. Kretinin}
\email{Present address:\\School of Physics and Astronomy\\ University of Manchester, Manchester, UK\\ e-mail: andrey.kretinin@manchester.ac.uk}
\affiliation{Weizmann Institute of Science, Condensed Matter Physics Department, Rehovot, Israel}
\author{Anindya Das}
\affiliation{Weizmann Institute of Science, Condensed Matter Physics Department, Rehovot, Israel}
\author{Hadas Shtrikman}
\affiliation{Weizmann Institute of Science, Condensed Matter Physics Department, Rehovot, Israel}

\begin{abstract}

\begin{enumerate}[I.]
\itemsep10pt
\item \textbf{CONTACT TRANSPARENCY}
\item \textbf{MAGNETIC AND TEMPERATURE DEPENDENCE OF \\THE CRITICAL CURRENT}
\item \textbf{SIMPLE MODEL OF THE JOSEPHSON JUNCTION COUPLED TO \\AN EXTERNAL RESONATOR}
\item \textbf{SPATIALLY AVERAGED STRAIN S IN THE CASE OF \\AN ASYMMETRIC NANOWIRE DEVICE}
\item \textbf{ANOTHER DEVICE USED IN THE EXPERIMENT}
\end{enumerate}

\end{abstract}

\maketitle

\section{Contact transparency}
\begin{figure}[!ht]
  \includegraphics{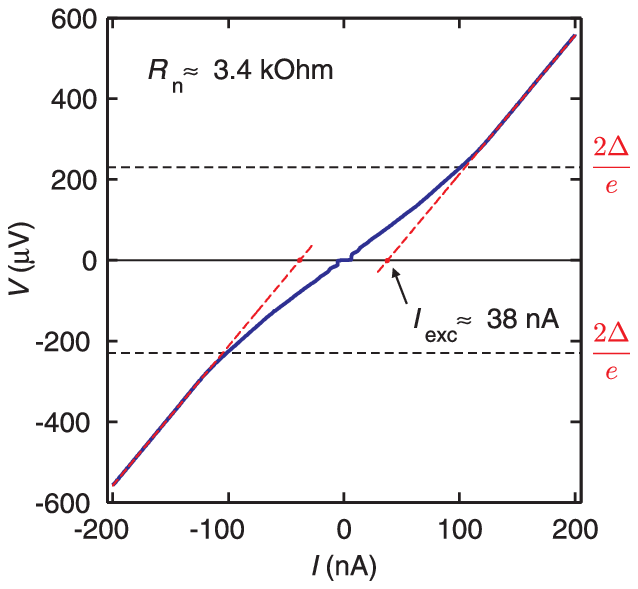}\\
  \caption{The excess current of the InAs nanowire Josephson junction described in the main text taken at $V_{\rm g}=$~12.3~V.}\label{S1}
\end{figure}
To assess the transparency of the superconducting Al contacts made to the InAs nanowire we measured the so-called excess current\cite{Blonder1982} $I_{\rm{exc}}$. The excess current of the Josephson junction is related to the presence of Andreev Reflections, and defined by the average strength $Z$ of the interface barrier between the normal weak-link and the superconducting contact. The value of $I_{\rm{exc}}$ can be determined as an intersection of the $I-V$ curve taken for $V>\pm2\Delta/e$, and extrapolated down to $V=$~0 as shown in Fig.~\ref{S1}. Using this method we found that $I_{\rm{exc}}\approx$~38~nA at $V_{\rm g}=$~12.3~V and the normalised excess current $\frac{R_{\rm n}I_{\rm{exc}}}{e\Delta}$ to be about 1.15, which corresponds to $Z\sim$~0.3 (see Fig.~6 in Ref.~\onlinecite{Blonder1982}). Hence, the contact transparency defined as $(1+Z^2)^{-1}$ is about 90$\%$.

\section{Magnetic and temperature dependence of the critical current}

\begin{figure}[!ht]
  \includegraphics{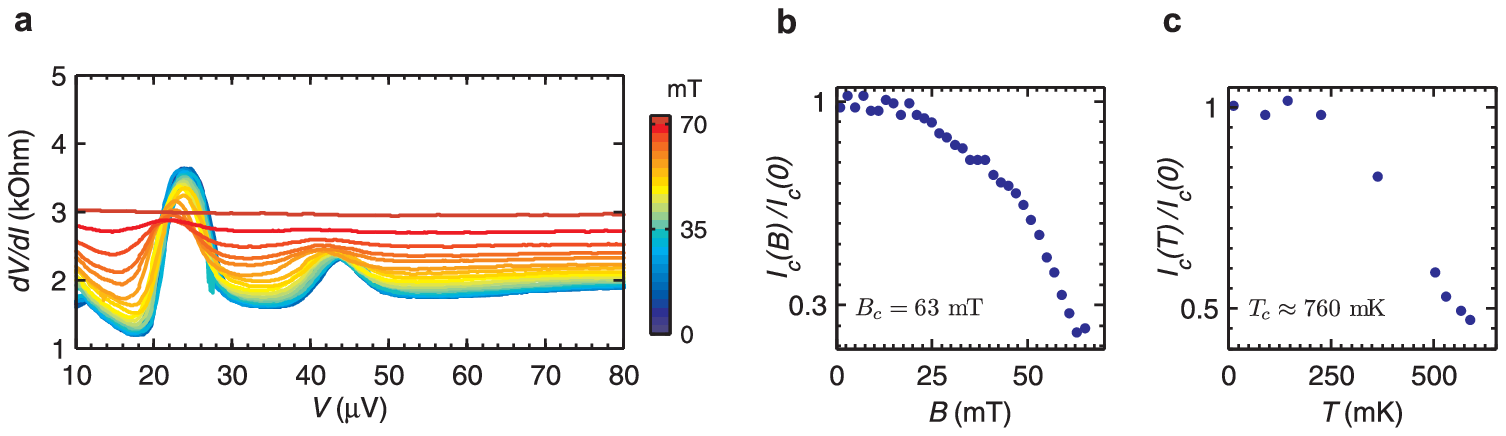}\\
  \caption{Additional data on the magnetic field and temperature dependence of $dV/dI$ and $I_{\rm c}$ for the nanowire device described in the main text.}\label{S2}
\end{figure}

Figure~\ref{S1}a shows the zoomed in plot of $dV/dI$ as a function of $V$ at different magnetic field shown in Fig.2a of the main text. The background resistance increasing with the magnetic field made it difficult to reliably extract the position of the resonances. However, it is clear from the plot that the resonance position is virtually independent of the magnetic field. Figure~\ref{S2}b is the plot of the critical current $I_{\rm c}$ as a function of the magnetic field $B$ measured at $T=T_{\rm base}$ and $V_{\rm g}=$~12.3~V, and Fig.~\ref{S2}c is the critical current $I_{\rm c}$ as a function of the temperature $T$ measured at $B=0$ and $V_{\rm g}=$~12.3~V.

\section{Simple model of the Josephson junction coupled to an external resonator}

\begin{figure}[!ht]
  \includegraphics{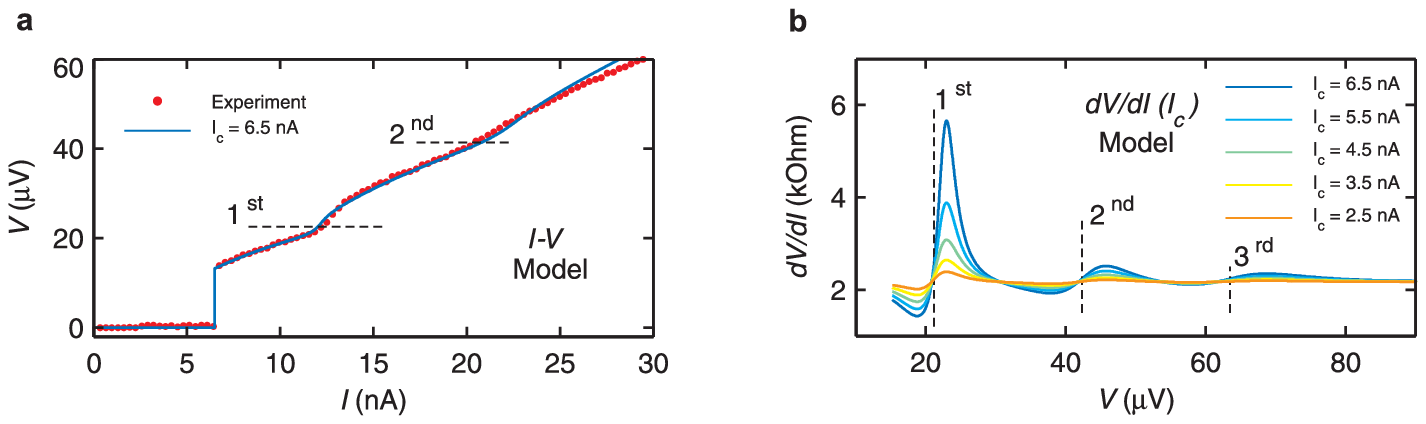}\\
  \caption{\textbf{Simple model of the underdumped Josephson junction coupled to an external electromagnetic resonator\cite{LikharevBook}.} \textbf{(a)} The theoretical $I-V$ curve (blue) and experimental data (red circles) taken at $T=T_{\rm base}$. \textbf{(b)} The theoretical differential resistance $dV/dI$ as a function of $V$ derived from the junction $I-V$ characteristics simulated for different values of $I_{\rm c}$. The plot clearly shows that the position of the resonance marked by vertical dashed lines corresponds to the intersection point of curved taken at different $I_{\rm c}.$}\label{S3}
\end{figure}

To verify the assumption made in the main text that the position of the resonance is given by the interception of the curves measured at different temperature we used a simple theoretical model of the underdumped Josephson junction weakly coupled to an external electromagnetic resonator.\cite{LikharevBook} Here the theoretical $I-V$ curve (Fig.~\ref{S3}a) is the superposition of the $I-V$ curve of the underdupmed Josephson junction $V = R_{\rm n}I$ and three current peaks $\delta I$ with Lorenzian lineshape $\delta I(V_{\rm{r}}) = 0.5zI_{\rm c}(1+\xi^2)^{-1}$, where $z\sim$~0.5 is the coupling coefficient, $\xi = 2Q(V/V_{\rm r}-1)$, and $Q\sim$~3 is the quality factor. These peaks are caused by the resonant coupling of the a.c.-Josephson oscillations to the external resonator characterised by the resonant voltage $V_{\rm r}$ and the quality factor $Q$. To account the effect of the temperature and magnetic field we assumed that the only way these parameters affect the $I-V$ curve is through changing critical current $I_{\rm c}$ as illustrated in Fig.~\ref{S2}b and Fig.~\ref{S2}c. Figure~\ref{S3}b shows the differential resistance $dV/dI$ derived from the theoretical $I-V$ curve. It is clear that the position of the resonance $V_{\rm r}$ corresponds to the interception point of the curves with different $I_{c}$, which confirms our assumption.

It is important to note that the Bulk Acoustic Resonator has two two distinct resonant frequencies.\cite{RosenbaumBook} The first one is at $\omega_{\rm r} = \pi\vartheta_{\rm a}/L$, where $Z(\omega_{\rm r})\rightarrow\infty$, which corresponds to the resonance of a parallel $LC$-circuit. Another one occurs at $\omega_{\rm r} = \vartheta_{\rm a}/L\left[\pi^{2}-8k^{2}\right]$, where $Z(\omega_{\rm r})\rightarrow0$, and it corresponds to the resonance of a series $LC$-circuit. The later one was not observed in our experiment, presumably, because the Josephson junction is already strongly underdumped by the relatively large value of $C_{\rm j}$, and the additional shunting of the a.c.-Josephson current does not take any visible effect on the $I-V$ curve.

\section{Spatially averaged strain $S$ in the case of an asymmetric nanowire device}

\begin{figure}[!ht]
  \includegraphics{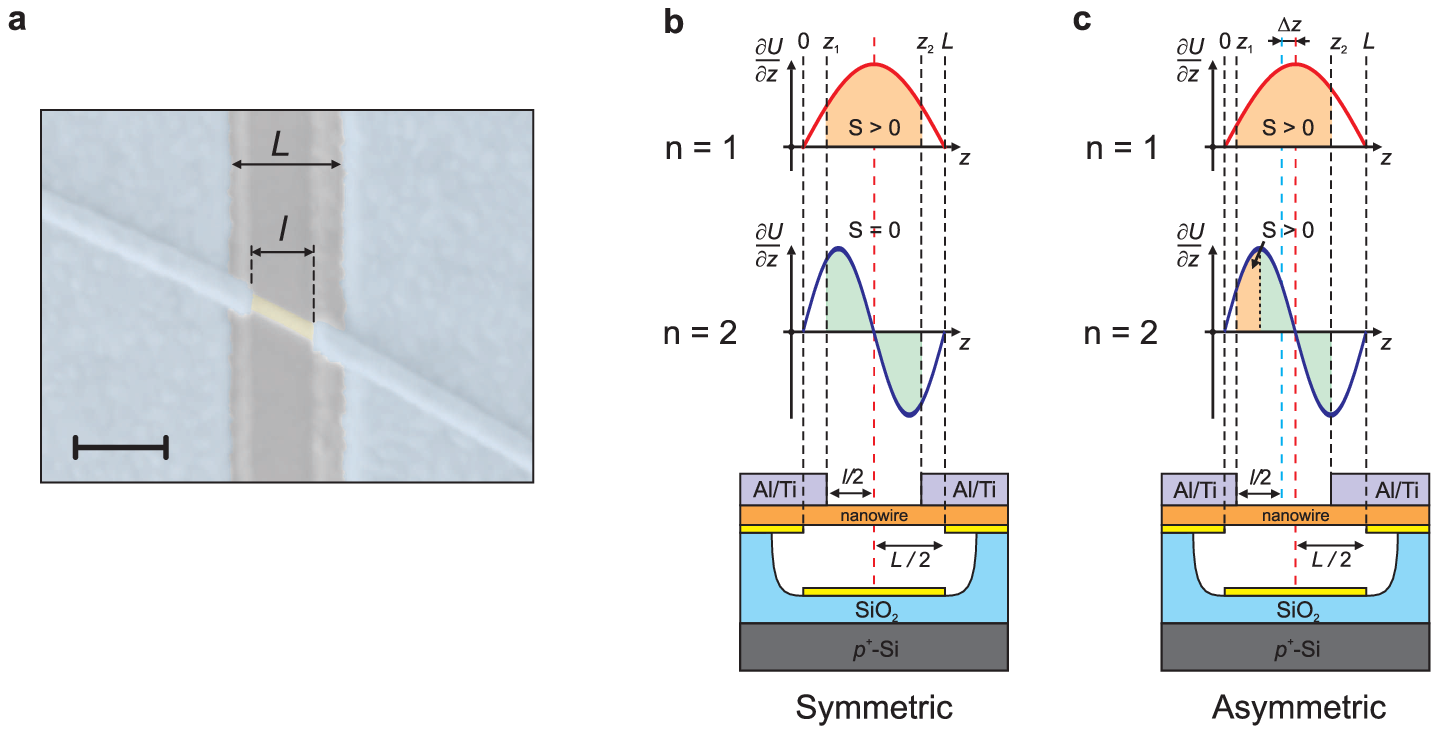}\\
  \caption{The asymmetric geometry of the experimental nanowire device and the non-zero spatially averaged strain $S$. \textbf{(a)} The scanning electron microscope image of a typical device, which shows the difference in mechanical $L$ and electrical $l$ length of the nanowire device. Also, the asymmetry due to the lithography misalignment is clearly visible. The scale bar corresponds to 200~nm. \textbf{(b)} The schematically represented device cross section and the mechanical strain distribution along the nanowire for the first two acoustic modes in the case of the symmetrically positioned contacts. The shaded areas under the strain distribution curves give the value of the integral (\ref{Eg.1}). \textbf{(c)} The same as \textbf{(b)}, but with the contacts being asymmetrically positioned with respect to the centre of the nanowire (shifted by $\Delta z$). The resulting non-zero values of the integral (\ref{Eg.1}) is shown for the case of $n=$~2.}\label{S4}
\end{figure}

As mentioned in the main text, the non-zero spatially averaged mechanical strain forbids the appearance of the even resonant modes of BAR. Indeed, the displacement current through the piezoelectric capacitor is $I_{\rm D} = C_{\rm j}[\dot{V}-(e_{33}/\epsilon)\dot{S}]$,\cite{Geller2005,Trees2005} where $e_{33}$ is the piezoelectric constant, $\epsilon$ is the dielectric constant, $S=\langle\partial U(z)/\partial z\rangle$ is the spatially averaged mechanical strain ($U(z)$ is displacement, $z$ is the coordinate along the piezoelectric media), and the dot denotes the time derivative. The second term in the expression is the piezoelectric contribution, which is non-zero when the media accommodates an odd standing acoustic wave. Assuming that the displacement $U(z)\propto \cos(k_{\rm n}z)$\cite{Geller2005}, where $k_{\rm n} = \pi n/L$, its spatial average is defined as
\begin{equation}
S=\left\langle\frac{\partial U(z)}{\partial z}\right\rangle=\frac{1}{L}\int_{0}^{\rm L}\frac{\partial U(z)}{\partial z} dz\propto\frac{1}{L}\int_{0}^{\rm L}\sin{\left(\frac{\pi nz}{L}\right)} dz,
\label{Eg.1}
\end{equation}
thus this integral is non-zero only for odd standing waves (odd values of $n$).

However, due to the self aligned geometry of our suspended nanowire device the distance $l$ between ohmic contacts is smaller than the distance $L$ between clumped nanowire ends and the position of their centres is slightly shifted with respect to each other (Fig.~\ref{S4}a). This shift ($\sim$25~nm) results from the finite accuracy of the e-beam lithography alignment. In this situation the averaging with integral~(\ref{Eg.1}) should take place only between the ohmic contacts along the length $l$ (from $z_{\rm 1}$ to $z_{\rm 2}$) as shown in Fig.~\ref{S4}b. For purely illustrative purpose we assume that $\partial U/\partial z(z=0,L)=$~0. For the symmetric arrangement of the contacts the first mode gives the non-zero value of $S$ and zero value for the second mode. In the asymmetric case (Fig.~\ref{S4}c), the first mode still gives the non-zero $S$. Apart from that, the difference $\Delta z$ between the position of the actual nanowire centre and the centre of the contacted segment makes the spatially averaged strain for the second mode non-zero as well. A similar approach is used to activate even acoustic resonances in the so-called Composite Bulk Acoustic Resonators.\cite{Rosenbaum1986,Zhang2003,Pang2010}

\begin{figure}[!ht]
  \includegraphics{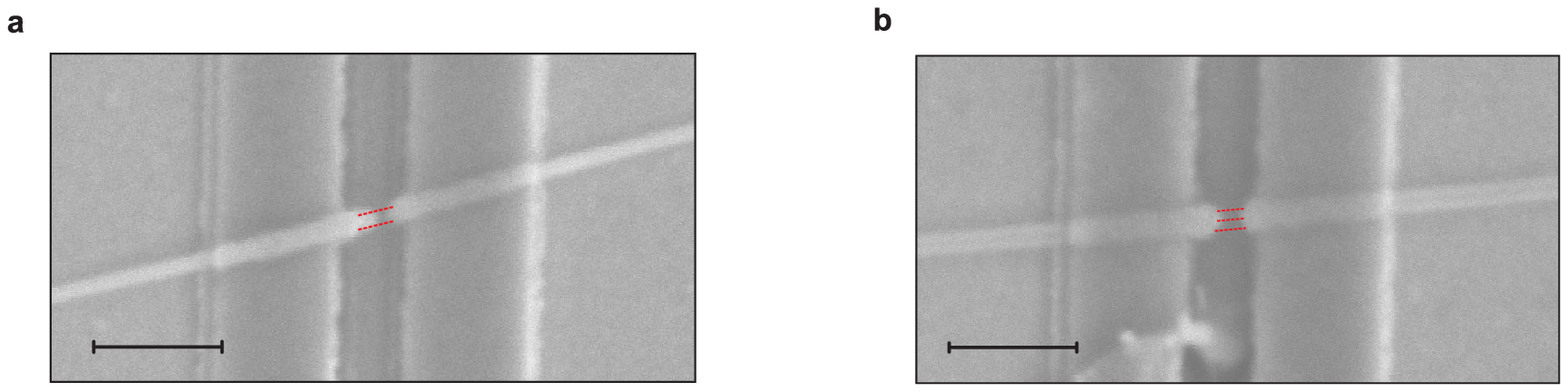}\\
  \caption{The scanning electron microscope images of the device described in the main text \textbf{(a)}, and additional device described here \text{(b)}. The scale bar corresponds to 500~nm. During the preparation for the SEM imaging both of the devices were broken down by an unintentional electrostatic discharge. The missing segments of the nanowire are marked by the red lines.}\label{S5}
\end{figure}

\section{Another device used in the experiment}

Apart from the device described in the main text and shown in Fig.~\ref{S5}a we also measured a more complex nanowire configuration, namely two thinner (30~nm in diameter) InAs nanowires held to each other side by side by the Van der Waals forces as shown in Fig.~\ref{S5}b. Both of the nanowires were suspended across the predefined trench and contacted in parallel with Ti/Al contacts, so they form a single weak link of the Josephson junction.

\begin{figure}[!ht]
  \includegraphics{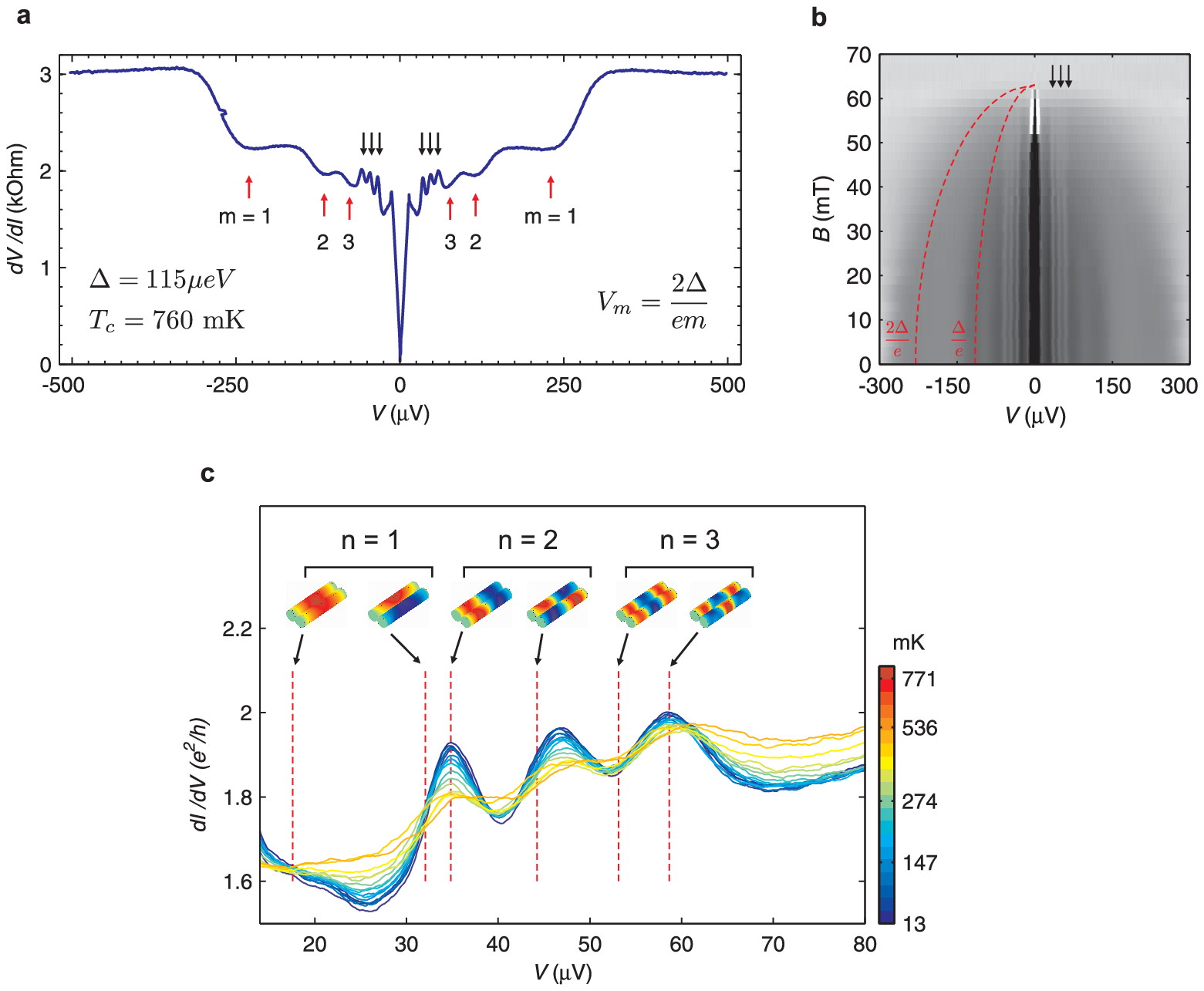}\\
  \caption{Additional experimental data obtained from a double beam nanowire device shown in Fig.~\ref{S5}b. \textbf{(a)} The differential resistance $dV/dI$ as a function of the voltage $V$ taken at $V_{\rm g}=$~0~V and $T=T_{\rm base}$. The red arrows show first three orders of MAR at voltage $V_{\rm m}$. The position of the first MAR signature ($m=$~1) gives the superconducting gap $\Delta\approx$~115~$\mu$eV and the estimate for $T_{\rm c}=2\Delta/(3.52\,k_{\rm B})\approx$~760~mK. The black arrows mark the resistance resonances caused by the acoustic wave resonance in double beam system. \textbf{(b)} The gray-scale plot of $dV/dI$ made in the $V-B$ plane at $T = T_{\rm base}$. The red dashed curves show the position of the first two orders of MAR expected from the Landau-Ginzburg theory,\cite{Douglass1961} $\Delta(B)/\Delta(0) = \sqrt{1-(B/B_{\rm c})^2}$, where $B_{\rm c}=$~63~mT is the critical field. The black arrows mark the magnetic field-independent position of the resistance resonances. The position of the resistance resonances stays unchanged while the peak amplitude decreases while $B$ approaches its critical values $B_{\rm c}$. \textbf{(c)} The zoomed-in plot of the $dV/dI$ resonance taken at different temperatures and zero magnetic field. Identically to the magnetic field dependence shown in \textbf{(b)}, the position of the resonances is independent of the temperature. The vertical dashed lines mark the voltage which corresponds to the first three longitudinal harmonics expected for the double InAs beam system with length 230~nm, diameter 30~nm and speed of sound $\vartheta_{\rm a}=~4.5\cdot10^3$~m/s. The coloured scale bar refers only to the $dV/dI$ traces taken at different temperature. \textbf{(Inset)} The schematically shown submodes of double beam system. The colour scale corresponds to the value of the spatially distributed strain inside each of the beam (red is maximum, blue is minimum). It is clear that the peak in the resistance appears to be closer to the resonance of the submode given by the antisymmetric combination of the main mode.}\label{S6}
\end{figure}

The differential resistance $dV/dI$ as a function of the bias $V$ taken at $V_{\rm g}=$~0~V and $T=T_{\rm base}$ is shown in Fig.~\ref{S6}a. As can be seen from the plot, apart from the MAR signatures with $m=$~1,~2,~3 (red arrows in Fig.~\ref{S6}a) the $dV/dI$ data also shows three closely spaced peak-like features in the range of $V$ between 30~$\mu$V and 70~$\mu$V (black arrows in Fig.~\ref{S6}a). Similarly to the single nanowire case described in the main text, the position of these features is independent of magnetic field and temperature as shown in Fig.~\ref{S6}b and \ref{S6}c. To evaluate the frequency of mechanical resonances of such double beam system we used numerical simulation made with COMSOL Multiphysics Package. We found that every acoustic mode turns out to be splitted into two submodes each of which is a symmetric or antisymmetric linear combination of a single beam acoustic mode (inset in Fig.~\ref{S6}c). Comparing the obtained frequencies with the position of the peaks in $dV/dI$ we concluded that their position corresponds to the antisymmetric submodes for $n=$~1,~2,~3. However, a more detailed theoretical study is needed to understand why only antisymmetric combination of the main modes are visible in the experiment.